\documentclass[twocolumn]{aastex6}

\slugcomment{Manuscript Version 3.1 on 28 Nov 2016}

\shorttitle{Water Ice in Edge-on Disks around Herbig Ae Stars}
\shortauthors{Terada \& Tokunaga}

\begin{document}

\title{Multi-Epoch Detections of Water Ice Absorption in Edge-on Disks around Herbig Ae Stars: PDS 144N and PDS 453}

\author{Hiroshi Terada}
\affil{Thirty Meter Telescope Project, National Astronomical Observatory of Japan, 100 West Walnut, Suite 300, Pasadena, CA 91124, USA}

\and

\author{Alan T. Tokunaga}
\affil{Institute for Astronomy, University of Hawaii, 2680 Woodlawn Drive, Honolulu 96822, USA}


\begin{abstract}
We report the multi-epoch detections of the water ice in 2.8--4.2\,$\micron$ spectra of two Herbig Ae stars, PDS 144N (A2 IVe) and PDS 453 (F2 Ve), which have an edge-on circumstellar disk. The detected water ice absorption is found to originate from their protoplanetary disks. The spectra show a relatively shallow absorption of water ice around 3.1\,$\micron$ for both objects. The optical depths of the water ice absorption are $\sim$0.1 and $\sim$0.2 for PDS 144N and PDS 453, respectively. Compared to the water ice previously detected in low-mass young stellar objects with an edge-on disk with a similar inclination angle, these optical depths are significantly lower. It suggests that stronger UV radiation from the central stars effectively decreases the water ice abundance around the Herbig Ae stars through photodesorption. The water ice absorption in PDS 453 shows a possible variation of the feature among the six observing epochs. This variation could be due to a change of absorption materials passing through our line-of-sight to the central star. 
The overall profile of the water ice absorption in PDS 453 is quite similar to the one previously reported in d216-0939 and this unique profile may be seen only at a high inclination angle in the range of 76--80$\arcdeg$.

\end{abstract}

\section{Introduction} \label{sec:intro}

Water is a crucial constituent for the formation of life and therefore how water is delivered to a protoplanet is of great interest. Water ice from the natal cloud is thought to accrete onto a protoplanetary core forming icy planetesimals and comets as a possible carrier of the water delivery\deleted{onto a protoplanetary core}. While the evolution of the water ice in the protoplanetary environments is important, astronomical detections for water ice in the protoplanetary disks are still too limited to develop a unified view of the water ice evolution in the disks \citep[][]{pon05,ter07,ter12a,ter12b,hon09,hon16}.

Edge-on disks are suitable for investigation of protoplanetary disks because the disk geometry is well defined with the central star occulted by the circumstellar disk. Recent sub-arcsecond resolution imaging with $Hubble$ $Space$ $Telescope$ and ground-based adaptive optics (AO) facilities enables us to spatially resolve disk formation sites, and the number of objects associated with a circumstellar disk morphology has been increasing. \citet{per06} discovered an edge-on disk around the intermediate-mass young stellar object, Herbig Ae star, PDS 144N. The disk around PDS 144N exhibits almost a perfect edge-on morphology and its inclination angle is estimated to be 83$\arcdeg$ \citep[][]{per06}. PDS 144N has a widely separated companion object, PDS 144S, which is also a Herbig Ae star \citep[][]{hor12}. PDS 453 is another Herbig Ae object showing an edge-on disk morphology with less inclination angle of 79$\arcdeg$ \citep[][]{per10}. These two edge-on disks are unique objects to investigate protoplanetary disks around higher mass YSOs since most of the detected edge-on disks are around low-mass YSOs. 

Water ice shows a strong absorption at 3\,$\micron$ with rich features to provide information about its grain size, crystallinity, and mixtures with the other ices \citep[][]{boo15}. Ground-based spectroscopy at 2.8--4.2\,$\micron$ is a powerful tool to reveal the water ice in the protoplanetary disks. In particular, wavefront correction at $\ge$3\,$\micron$ is relatively easy and AO-assisted spectroscopy is very beneficial not only for high spatial observation but also for obtaining stable and reliable spectra. In this wavelength region, thermal background from the ambient background is still acceptably low and this allows achieving a high signal-to-noise ratio.

In Section \ref{sec:obs}, methods for observations and data analysis are described. Section \ref{sec:res} shows the results of the 3\,$\micron$ water ice absorption features for PDS 144N, PDS 144S, and PDS 453. Characteristics of the detected water ice absorption profiles are discussed in Section \ref{sec:dis}, including its possible variability in time and similarity to the profile of the edge-on disk object d216-0939. The conclusion is summarized in Section \ref{sec:sum}.

\section{Observation \& Data Reduction} \label{sec:obs}

All the observations were performed using the Infrared Camera and Spectrograph \citep[IRCS;] []{tok98,kob00} mounted on the Nasmyth platform of the Subaru Telescope. The Subaru AO system \citep[AO188;][]{hay10} was utilized only for the observations of PDS 453. The object itself served as a wavefront correction reference, since it is relatively bright in the optical ($R$$\sim$12.5). Imaging of PDS 453 at $K^{\prime}$($\lambda$=2.12\,$\micron$) and $L^{\prime}$($\lambda$=3.77\,$\micron$) was conducted with 20 mas camera of the IRCS on a single epoch of 2009 June 3, and spectroscopy at 2.8--4.2\,$\micron$ is carried out in two and six epochs from 2006 to 2014 for PDS 144N and PDS 453, respectively.  The airmass mismatch between spectroscopy of the object and the standard star was minimized to avoid any artificial features in the spectra due to mismatch of the telluric absorption and the resultant airmass difference was $\sim$0.01. In the second epoch of spectroscopy for PDS 144N, no standard star was observed and PDS 144S acts as a spectral standard star for the exact cancellation of the telluric absorption to obtain the PDS 144N spectrum. Sky conditions were excellent for all the observing epochs in terms of its extinction, seeing, and precipitable water level.  The observation details are summarized in Table~\ref{tbl-obslog}.

Five and nine-point box-shaped dithering patterns were used for imaging of PDS 453 at $K^{\prime}$ and $L^{\prime}$ with a separation of 3\arcsec, respectively. For spectroscopy of PDS 453, the slit was set with a position angle of 133$\arcdeg$ to align to the scattered light disk only for the first epoch observation on 2009 August 17. In the other observing epochs, the slit position angle was 0$\arcdeg$. The slit position angle for PDS 144N was 119$\arcdeg$ in order to locate PDS 144S on the same slit. A-BB-A nodding operation was performed for all the spectroscopy. Nodding separation for PDS 453 was 3$\arcsec$ for the first epoch and 1.5$\arcsec$ for the other epochs. In case of PDS 144N, the nodding separation of 2.5$\arcsec$ and 3.0$\arcsec$ was chosen for the first and the second epochs respectively to avoid interference with PDS 144S located with a separation of 5\farcs40. In spectroscopy of PDS 144N under natural seeing conditions, the slit widths were selected to be 0\farcs6 (corresponding spectral resolving power, R$\sim$190) in the first epoch and 0\farcs45 (R$\sim$260) in the second epoch. For AO spectroscopy of PDS 453, it was 0\farcs225 (R$\sim$510) throughout all the epochs. 

Data reduction was performed by IRAF software packages through a standard procedure that consists of flat fielding, sky subtraction, telluric correction, and wavelength calibration. HIP 79229 \citep[A0V, $V$=6.64 mag;][]{hog00} was used with an assumption of $V$$-$$L^{\prime}$=0 for a photometric estimate of PDS 453 at $L^{\prime}$. No $K^{\prime}$ photometric calibration was performed, because the peak signal in the image of PDS 453 at $K^{\prime}$ was saturated,  For spectroscopic reference, four A0 stars, HR 5197, HR 6061, HR 6354, and HR 6490 were observed for correction of the telluric absorption. The hydrogen absorption feature from the A0 stars were removed using the method by \citet{vac03}. The telluric absorption lines in the spectrum was used for the wavelength calibration. 

\begin{deluxetable*}{l l c c c c c c}
	\tabletypesize{\scriptsize}
    \tablecolumns{8}
    \tablewidth{0pt}
    \tablecaption{Observing log\label{tbl-obslog}}
    \tablehead{
      \colhead{}&\colhead{Observing}&\colhead{}&\colhead{Exposure}&\colhead{Spectral}&\colhead{Average}&\colhead{Standard}&\colhead{$\Delta$Airmass}\\
      \colhead{Object}&\colhead{Date}&\colhead{Mode}&\colhead{Time}&\colhead{Resolution}&\colhead{Airmass}&\colhead{Star}&\colhead{Obj.-Std.}\\
      \colhead{}&\colhead{(UT)}&\colhead{}&\colhead{(s)}&\colhead{($\lambda$/$\Delta\lambda$)}&\colhead{}&\colhead{}&\colhead{}}
      \startdata
      PDS 144S \& N&2006 Feb 7&$L$ Spectroscopy&1080&190&1.497&HR 5197&$-$0.025\\
      &2008 Feb 18&$L$ Spectroscopy  &840&260&1.470&\nodata&\nodata\\
      PDS 453&2009 Jun 3&$K^{\prime}$ Imaging&75&\nodata&&\nodata&\nodata\\
	&                      &$L^{\prime}$ Imaging&180&\nodata&1.495&HIP 79229&$-$0.007\\
      &2009 Aug 17&$L$ Spectroscopy  &480&510&1.467&HR 6061&$+$0.001\\
      &2011 Aug 14&$L$ Spectroscopy  &360&510&1.531&HR 6490&$-$0.010\\
      &2011 Aug 19&$L$ Spectroscopy  &480&510&1.572&HR 6490&$-$0.014\\
      &2012 Sep 17&$L$ Spectroscopy  &480&510&1.594&HR 6354&$+$0.0002\\
      &2014 Mar 20&$L$ Spectroscopy  &720&510&1.440&HR 6490&$+$0.008\\
      &2014 Mar 21&$L$ Spectroscopy  &720&510&1.439&HR 6490&$-$0.001\\
      \enddata
\end{deluxetable*}


\section{Results} \label{sec:res}
We present the 2.8--4.2\,$\micron$ spectra for two Herbig Ae stars (PDS 144N and PDS 453) with an edge-on morphology of the surrounding disks and extract the water ice absorption feature in the spectrum. For extraction of the feature, the continuum of the spectrum was estimated using a second-order polynomial fitting with wavelength regions of 2.875--2.89\,$\micron$ and 3.7--4.0\,\micron. Since the water ice profile is known to continue in the wavelength region of $<$2.88\,\micron, it is noted that this continuum determination could cause an underestimate of the water ice absorption by $\sim$25\% at the peak. 

\subsection{PDS 144N}
\begin{figure}
 \begin{center}
   \plotone{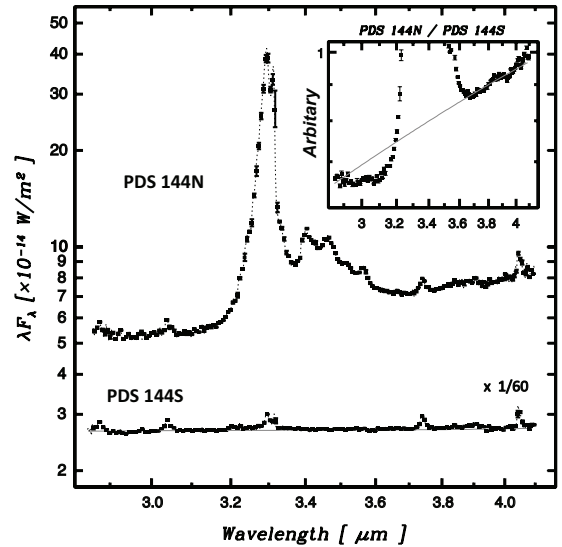}
   \caption{\label{fig-pds144-1st}Spectra of PDS 144N and PDS 144S. The inset shows a spectrum of PDS 144N simply divided by PDS 144S, which is normalized at 4.1\,$\micron$. The estimated continuum is shown by the grey solid lines. While PDS 144S exhibits no absorption feature, a shallow absorption around 3.1\,$\micron$ is seen in the spectrum of PDS 144N.}
 \end{center}
\end{figure}

PDS 144N exhibits a clear edge-on morphology with a high spatial resolution imaging, while its binary companion, PDS 144S shows no apparent disk structure \citep{per06}. Spectroscopy for PDS 144N and PDS 144S were simultaneously done with an appropriate position angle (119$\arcdeg$) of the slit. The spectra of both objects are shown in Figure~\ref{fig-pds144-1st} for the first epoch (2006-02-07-UT) observation. The absolute flux is calibrated with $K$ and $L^{\prime}$ magnitudes derived by \citet{per06}. The most prominent feature in the spectrum of PDS 144N is a PAH emission feature ranging from 3.2\,$\micron$ to 3.6\,\micron, and the shallow water ice absorption feature is seen in the wavelength region of 2.9--3.2\,\micron. On the other hand, the spectrum of PDS 144S is featureless ($\tau_{ice}$$\le$0.002) and flat in this low resolution spectrum except for the hydrogen recombination lines at 2.873\,\micron, 3.039\,\micron, 3.297\,\micron, 3.741\,\micron, and 4.052\,\micron, which means that PDS 144S can act as an atmospheric calibrator to obtain nearly perfect cancellation of the telluric absorption. 

\begin{figure}
 \begin{center}
   \plotone{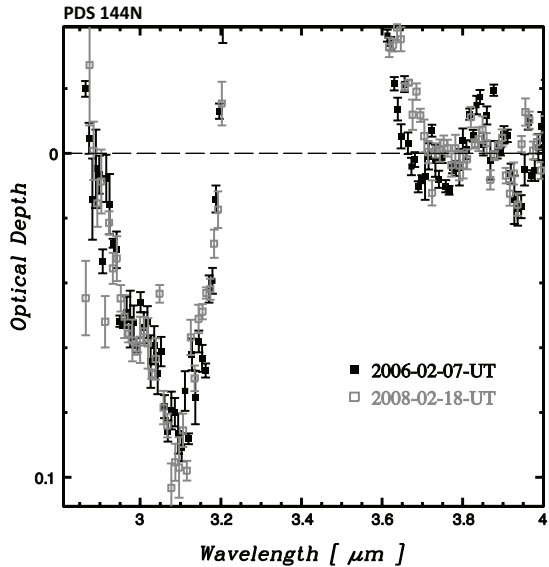}
   \caption{\label{fig-pds144-tau}Water ice optical depth of PDS 144N in the first (2006-02-07-UT) and second (2008-02-18-UT) epochs showing good agreement.
   }
 \end{center}
\end{figure}

After extraction of the water ice absorption from the spectrum of PDS 144N in the first and second (2008-02-17-UT) epochs using PDS 144S for the telluric absorption cancellation, their optical depths are plotted in Figure~\ref{fig-pds144-tau}. There is no significant change of the optical depth ($\tau_{ice}$=0.09$\pm$0.01) and the wavelength of maximum optical depth ($\sim$3.09\,$\micron$) in the two epochs.

\subsection{PDS 453}
\begin{figure}
 \begin{center}
   \plotone{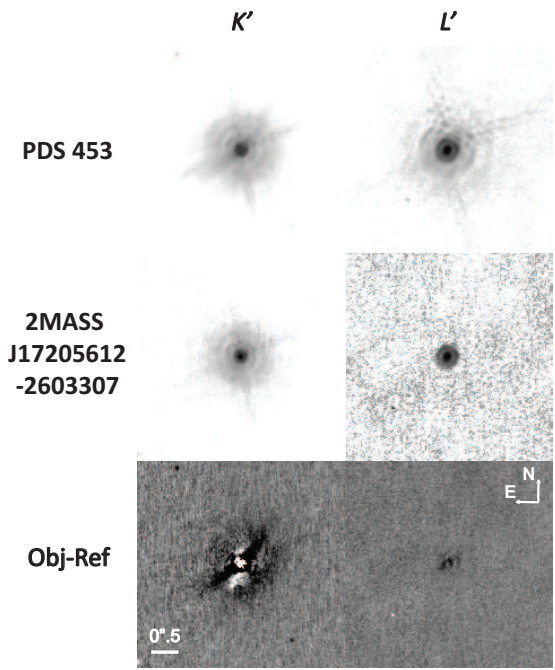}
   \caption{\label{fig-pds453-img}Adaptive Optics Images of PDS 453 at $K^{\prime}$ and $L^{\prime}$. The object and the nearby star as PSF reference are shown at the top and the middle panels, respectively. The residual signals of the PSF subtracted images are seen in the bottom panel. The PSF subtracted image at $K^{\prime}$ clearly shows the nearly edge-on morphology of the circumstellar disk around PDS 453 at a position angle of 133$\arcdeg$.}
 \end{center}
\end{figure}

Figure~\ref{fig-pds453-img} demonstrates the AO images taken at $K^{\prime}$ and $L^{\prime}$.  Aperture photometry is applied to the $L^{\prime}$ image with a radius of 1\farcs5 to find $L^{\prime}= 8.10$ mag. The photometric error is $\sim$0.1 mag. After subtracting the object images by the normalized images of the nearby star (2MASS J17205612-2603307), the residual images are displayed at the bottom panel. Scattered light disk around the object can be seen in the point-spread-function (PSF) subtracted image at $K^{\prime}$, which is consistent with the discovery result of \citet{per10}. At $L^{\prime}$, no significant structure is found in the PSF subtracted image. The normalization factor for $L^{\prime}$ is determined taking into account their brightness at $L^{\prime}$. For the $K^{\prime}$ image,  the factor is searched to obtain the best subtraction of a speckle pattern around the object. Since the nearby star is fainter than the object, the signal-to-noise ratio of the PSF is limited by the nearby star. 

\begin{figure}
 \begin{center}
   \plotone{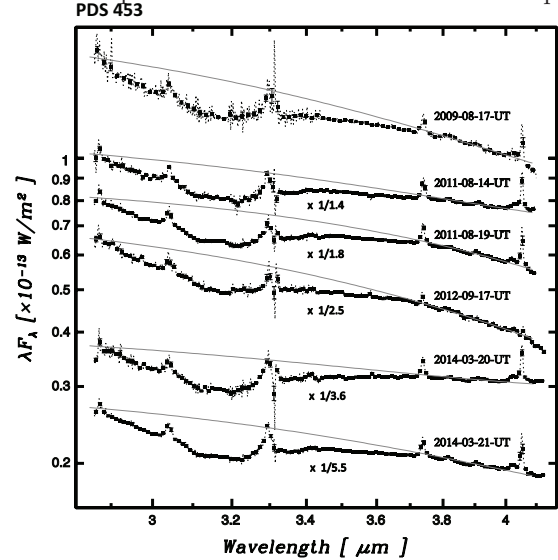}
   \caption{\label{fig-pds453-spec}Spectra of PDS 453 at six epochs. Each spectrum is offset for a better presentation. Solid grey lines are the estimated continuum with a second-order polynominal function.
   }
 \end{center}
\end{figure}

\begin{figure}
 \begin{center}
   \plotone{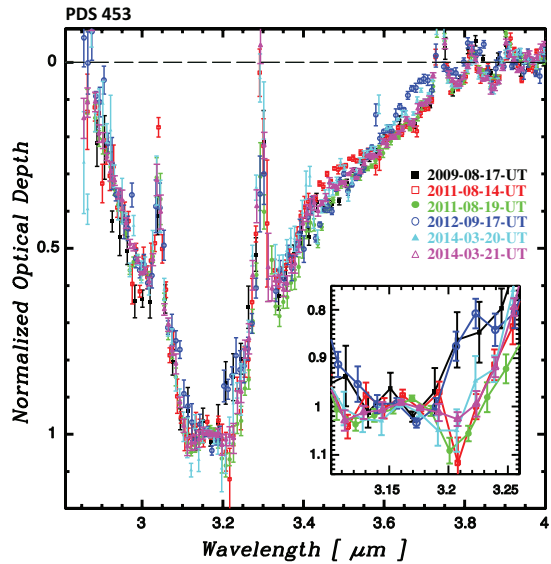}
   \caption{\label{fig-pds453-var} Normalized optical depth of the detected water ice in PDS 453 in six observing epochs. The inset shows a close-up view around the peak depths to show the possible variation of the profile in 3.2--3.25\,$\micron$.
   }
 \end{center}
\end{figure}

The 2.8--4.2\,$\micron$ spectra are presented in Figure~\ref{fig-pds453-spec} for 6 epochs from 2009-08-17-UT to 2014-03-21-UT, after being normalized for the observed $L^{\prime}$ magnitude of 8.10. The slope of the continuum was changing with time, and a shallow water ice absorption was clearly detected in all the spectra. 
In all the epochs, the water ice absorption has a depth of 0.19$\pm$0.01 with a wide absorption band ranging from 3.10\,$\micron$ to 3.23\,$\micron$.
To look into the water ice profile, their normalized optical depths are presented in Figure~\ref{fig-pds453-var}. The normalization is applied to the averaged optical depth in the wavelength range of 3.15--3.18\,\micron, where it is almost free from the telluric absorption. While the overall profile is quite consistent throughout six epochs, an apparent change of the water ice absorption profile can be seen at 3.20--3.25\,$\micron$. More specifically, data on 2009-08-17-UT and 2012-09-17-UT shows deviations from the other spectra. 
This change of the water ice absorption profile is discussed in the Section \ref{sec:dis}.

\section{Discussion} \label{sec:dis}

Physical parameters for PDS 144N, PDS 144S, and PDS 453 are summarized in Table~\ref{tbl-tarpara}. Although the distance to PDS 144N was originally suggested to be around 1000pc \citep{per06}, more recent investigation favors the smaller value of 145pc \citep{hor12}. The distance to PDS 453 is more uncertain, but the value of 140pc assumed by \citet{per10} is adopted here.
\begin{deluxetable*}{l c c c c c c}
	\tabletypesize{\scriptsize}
    \tablecolumns{7}
    \tablewidth{0pt}
    \tablecaption{Target Parameters\label{tbl-tarpara}}
    \tablehead{
      \colhead{}&\colhead{Spectral}&\colhead{Inclination}&\colhead{Disk}&\colhead{Possible}&\colhead{}&\colhead{}\\
      \colhead{Object}&\colhead{Type}&\colhead{Angle}&\colhead{Diameter}&\colhead{Association}&\colhead{Distance}&\colhead{Reference}\\
      \colhead{}&\colhead{}&\colhead{($\arcdeg$)}&\colhead{($\arcsec$)}&\colhead{}&\colhead{(pc)}&\colhead{}}
      \startdata
      PDS 144N&A2IV&83$\pm$1&0.8&Upper Scorpius&145$\pm$2&\citet{per06}, \citet{hor12}\\
      PDS 144S&A5V&73$\pm$7 &0.8&Upper Scorpius&145$\pm$2&\citet{per06}, \citet{hor12}\\
      PDS 453&F2V&79$\pm$3&3.1&Scorpius-Centaurus&140&\citet{per10}\\
      \enddata
\end{deluxetable*}

\subsection{Location of Detected Water Ice}

Water ice absorption towards young stellar objects is often attributed to the foreground cloud materials residing in front of the targets, and that possibility is investigated here. 

While PDS 144S shows no water ice absorption, its binary Herbig Ae star, PDS 144N exhibits a shallow water ice absorption around 3.1\,$\micron$. Therefore, the detected water ice toward PDS 144N is confirmed to be localized around PDS 144N with a radius of less than 5\farcs40 (783au) and it is most likely attributed to the circumstellar protoplanetary disk of PDS 144N. 

\begin{figure}
 \begin{center}
   \plotone{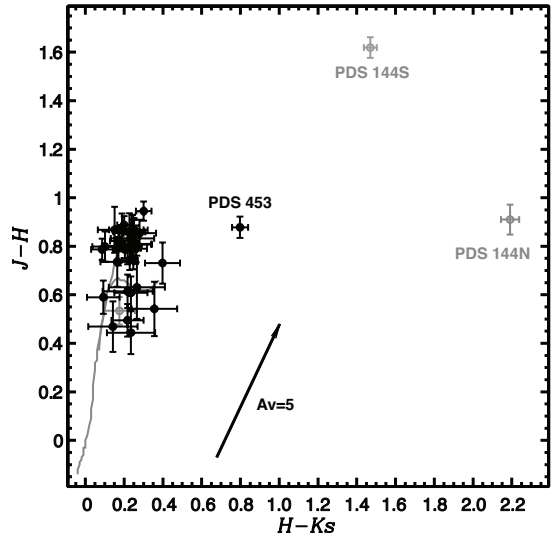}
   \caption{\label{fig-colcol}Two-color diagram for PDS 453 field (71\farcs4 $\times$ 71\farcs4). The gray lines indicates a location of dwarfs and giants.  The arrow shows an extinction vector of $A_{V}=5$ as a reference. Only PDS 453 exhibits a red color and it is separated significantly from the others. This indicates that the circumstellar material is localized around PDS 453. Just for reference, the PDS 144 field (69\farcs0 $\times$ 69\farcs0; 10000au $\times$ 10000au) is also presented with a gray open circles, in which PDS 144N and PDS 144S are found to be much redder than PDS 453. 
   }
 \end{center}
\end{figure}

Regarding PDS 453, there is no bright nearby star around the object to use as a comparison. 
To see the surrounding environment around PDS 453 within 71\farcs4 $\times$ 71\farcs4 (corresponding to 10000au $\times$ 10000au), Figure~\ref{fig-colcol} shows 2-color diagram by using $J$, $H$, and $K_{s}$ photometry with the 2MASS catalog . In this figure, it is seen that PDS 453 is separated significantly in these $J$-$H$ and $H$-$K_{s}$ colors and therefore the absorbing material is localized around PDS 453.

\begin{figure}
 \begin{center}
   \plotone{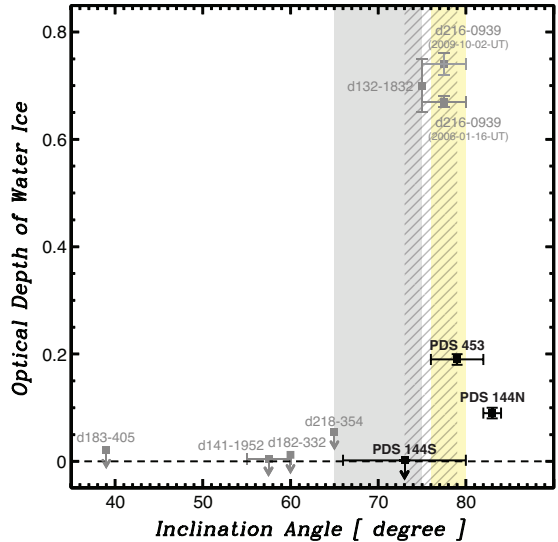}
   \caption{\label{fig-inctau} Water ice optical depth as a function of the inclination angle of the disks. Black squares are for the edge-on disks around Herbig Ae stars. Gray squares show optical depths of the water ice detected in the silhouette disks of the Orion nebula cluster and M43 taken from Figure 9 of \citet{ter12b}. For clarity, data points for d121-1925 and d053-717 are not shown because the detection toward d121-1925 is attributed to the foreground ice and d053-717 is suspected to be not in the silhouette disk. The shaded area correcponds to the critical inclination angle range suggested by \citet{ter12b} for disks around the low-mass YSOs. In this figure, the critical inclination angle to show the water ice in the Herbig Ae disks is found to be 73--79$\arcdeg$, which is shown with the diagonal lines. Area in yellow shows a key inclination angle (76--80$\arcdeg$) to exhibit the wider water ice absorption in protoplanetary disks shown in Figure~\ref{fig-pds453-comp} and discussed in Section \ref{sec:sim}.
      }
 \end{center}
\end{figure}

According to \citet{hor12}, the inclination angle of the circumstellar disk around PDS 144S is very high (73$\pm$7\arcdeg). On the other hand, no signature for the scattered light morphology of the disk is seen in the infrared image of PDS 144S, which implies a small inclination angle for the circumstellar disk. \citet{ter12b} found a threshold of the inclination angle of 65--75$\arcdeg$ for protoplanetary disks that exhibit the water ice absorption in the low-mass young stellar objects in the Orion nebula cluster and M43 regions. In analogy to the result for low-mass YSOs, no detection of water ice absorption towards PDS 144S suggests a critical inclination angle of more than 73$\arcdeg$ for the detection of water ice absorption on the assumption that PDS 144S has similar abundance of the water ice in the circumstellar disk as PDS 144N and PDS 453. Figure~\ref{fig-inctau} shows the optical depth of the water ice detected for PDS 144S, PDS 144N, and PDS 453 together with data from \citet{ter12b}. Here, the critical inclination angle for showing the water ice absorption appears to be 73--79$\arcdeg$ for the disks around the intermediate-mass YSOs.

\subsection{Effect of UV Radiation from the Herbig Ae Stars on Water Ice in the Disks}

Water ice in the circumstellar disk is known to be affected through a photodesorption process by the far UV (FUV) radiation \citep[e.g.,][]{oka12}. Since the far UV radiation from the central stars is supposed to be harsher in the Herbig Ae star system than in the low-mass young stellar object system, the water ice distribution especially at the disk surface could be totally different between these two systems. 

Due to the high quality spectra with a signal-to-noise of $\ge$100, very shallow water ice absorption could be detected with $\tau$$\sim$0.1 and 0.2 for PDS 144N and PDS453, respectively. These values are significantly smaller compared with that for the low-mass young stellar objects associated with edge-on disks ($\tau$$=$0.7--1.7), HK Tau B, HV Tau C, and d216-0939 \citep{ter07,ter12b}. This can be qualitatively explained by the stronger photodesorption process. 

Also for the possible larger critical angle (73--79\arcdeg) of the disk inclination for producing the water ice absorption, the photodesorption effect by the stronger UV radiation may be the primary cause. The stronger FUV radiation pushes the snow line at the disk surface further into the disk mid-plane, and as a result the opening angle of the ice region in the disk will be smaller.

\subsection{Water Ice Absorption Profile and Similarity between PDS 453 and d216-0939} \label{sec:sim}

The strong PAH emission features seen in PDS 144N prevents from an accurate evaluation for the entire absorption profile of the water ice. However, PAH emission is considered to exhibit its feature typically from 3.2\,$\micron$ to 3.6\,$\micron$ \citep[e.g.,][]{tok91,vandie04,tie08}, and we assume a negligible contribution of the PAH emission to the optical depth of the water ice at wavelengths $\le$3.2\,\micron. The water ice profile of PDS 144N presented in Figure~\ref{fig-pds144-tau} shows a water ice absorption at a peak wavelength of $\sim$3.08\,$\micron$, which is similar to the one in the edge-on disks of the low-mass young stellar objects \citep{ter07}. 

\begin{figure}
 \begin{center}
   \plotone{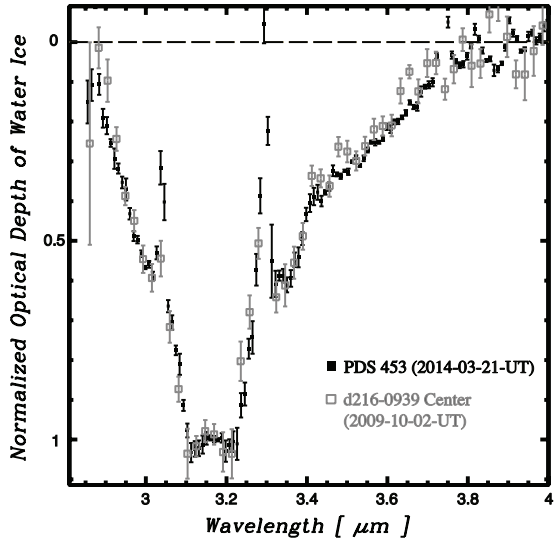}
   \caption{\label{fig-pds453-comp}Comparison between the water ice absorption profiles of PDS 453 and d216-0939. The best quality data is chosen among multi-epoch data set. Both the profiles are nearly identical around the peak of the optical depth.
   }
 \end{center}
\end{figure}

Regarding PDS 453, the overall profile of the detected water ice absorption exhibits an enhanced optical depth around 3.2\,$\micron$ (see Figure~\ref{fig-pds453-var}). It resembles the features reported for the silhouette disk object, d219-0939 in the M43 region \citep{ter12a}, in which the feature is interpreted as a large particle size ($\sim$0.8\,$\micron$) crystallized water ice absorption. The same procedure for extraction of the water ice feature was applied to the PDS 453 and d219-0939 data. The two optical depths are plotted in Figure~\ref{fig-pds453-comp} for the best quality spectra of PDS 453 (2014-03-21-UT) and d216-0939 (center position on 2009-10-02-UT)  with a normalization around 3.15--3.18\,$\micron$. It  clearly shows a very similar water ice absorption in these objects. This absorption feature is unique among the water ice absorption profiles detected so far in various kinds of astronomical targets \citep{boo15}. Taking into account that both PDS 453 and d219-0939 have similar inclination angles of their disks around 78$\pm$2$\arcdeg$ with a line-of-sight to the disk surface (see Figure~\ref{fig-inctau}), this suggests that a unique phenomena for grain growth and crystallization process goes on at the icy disk surface to produce this peculiar feature.

\subsection{Possible Time Variability of Water Ice Absorption}

As described in the Section \ref{sec:res}, the absorption profile of the water ice detected for PDS 453 at multiple epochs exhibits variability in the wavelength region of 3.2--3.25\,\micron, whereas the absorption feature of PDS 144N is found to be the same for the achieved signal-to-noise ratio. It is noted that the behavior of the absorption features in the 3.2--3.25\,$\micron$ is correlated with a feature at 2.97\,\micron. As seen in Figure~\ref{fig-pds453-spec}, the small absorption features in the 3.2--3.25\,$\micron$ wavelength region in the epochs of 2009-08-17-UT and 2012-09-17-UT are associated with the emission feature at 2.97\,\micron. In both the wavelength regions, strong telluric water vapor features exist and 
there is a possibility that this apparent variability is due to inappropriate estimate of the telluric absorption. 

\begin{figure}
 \begin{center}
   \plotone{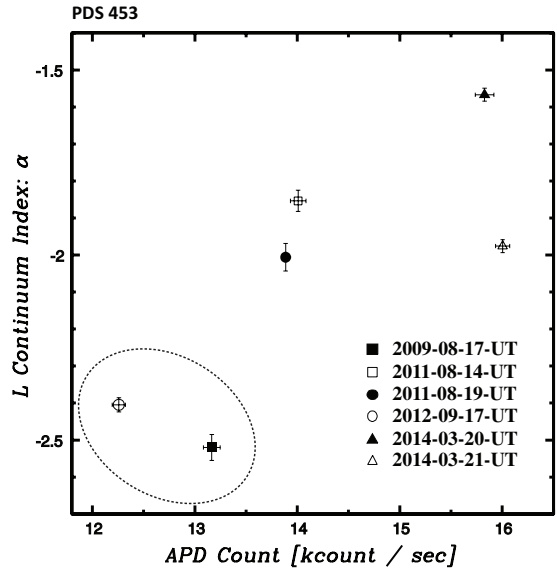}
   \caption{\label{fig-pds453-crr} APD counts of AO188 system vs. $L$ continuum slope index of PDS 453 spectra. 
   }
 \end{center}
\end{figure}

Even given that difficulty in identification of the variation source, it is still very interesting here to note the large photometric variation ($\Delta$$V$$\sim$1 mag) of PDS 453 in the optical found in the ASAS3 survey \citep{poj02}, which may suggest the variable extinction due to the absorbing material change in the sightline to the disk \citep{per10}. In fact, signals in the optical obtained by an abalanche photodiode (APD) on the AO188 system for the wavefront sensing with PDS 453 show a variation through these observation epochs. In addition, the continuum slope change is seen in the 3\,$\micron$ spectra. We
define the $L$ continuum slope index as $\alpha$=(log(I$_{\lambda_{2}}$)-log(I$_{\lambda_{1}}$))/(log($\lambda_{2}$)-log($\lambda_{1}$)), where $\lambda_{1}$=2.875--2.89\,$\micron$ and $\lambda_{2}$=3.7--4.0\,$\micron$, and the obtained APD counts and continuum index ($\alpha$) are plotted in Figure~\ref{fig-pds453-crr}. As shown with a dashed circle, both the APD count and the $\alpha$ index are located in the lower-left area for two epochs of 2009-08-17-UT and 2012-09-17-UT, in which the different feature of the water ice absorption profile is exhibited in the wavelength region of 3.20--3.25\,$\micron$. Although systematic errors are not taken into account for the APD count and the slope index, this correlation may imply the real change of the water ice absorption feature in PDS 453. 

\section{Summary} \label{sec:sum}

We summarize the results of this study as follows.
\begin{itemize}
  \item[1.]{Shallow 3\,$\micron$ water ice absorption features of two Herbig Ae stars with edge-on disks, PDS 144N and PDS 453, are detected. The absorption originates from the protoplanetary disks. }
  \item[2.]{No water ice absorption is detected towards PDS 144S. It indicates that the critical inclination angle to show the water ice absorption is larger in Herbig Ae disks than in low-mass young stellar disks. The larger critical inclination angle and shallower water ice absorption could be due to photodesorption of ice by the harsher FUV radiation from the Herbig Ae stars.}
  \item[3.]{The unusual profile of the water ice absorption detected in PDS 453 is very similar to the one found in d216-0939.  The observations suggest that an inclination angle of 76--80$\arcdeg$ is needed to show this feature which is attributed to larger ice grains with high crystallinity.}
  \item[4.]{Water ice absorption features detected in multi-epoch 2.8--4.2\,$\micron$ spectra of PDS 453 show a possible variation correlated with the $L$ continuum slope and optical brightness. It may be caused by variable absorption at the disk surface.}
\end{itemize}

\acknowledgements
We thank the entire support staff at the Subaru telescope for their efforts in  keeping this very complicated facility operational, in particular the instrument maintenance staff whose efforts kept the instrument stable and allowed us to obtain reliable monitoring observations over a long period of time. The authors wish to recognize and acknowledge the very significant cultural role and reverence that the summit of Mauna Kea has always had within the indigenous Hawaiian community.  We are most fortunate to have the opportunity to conduct observations from this mountain.

\facility{Subaru(IRCS, AO188)}

\end{document}